\documentclass[12pt,preprint]{aastex}

\shorttitle{Radio Confusion on Cluster Sunyaev-Zel'dovich Effect}
\shortauthors{Lin, Chiueh \& Wu}

\begin{document}

\title{Radio Confusion on Cluster Sunyaev-Zel'dovich Effect}

\author{Huichun Lin\altaffilmark{1}, Tzihong Chiueh\altaffilmark{2}}
\affil{Department of Physics, National Taiwan University, Taipei
106, Taiwan}

\and

\author{Xiang-Ping Wu\altaffilmark{3}}
\affil{National Astronomical Observatories, Chinese Academy of
Sciences, Beijing 100012, China}

\altaffiltext{1}{Email: deutsch@stella.astro.ncu.edu.tw}
\altaffiltext{2}{Adjunct Research Fellow, Institute of Astronomy
and Astrophysics, Academia Sinica, Taipei 106, Taiwan \\
Email: chiuehth@phys.ntu.edu.tw} \altaffiltext{3}{Email:
wxp@bao.ac.cn}

\begin{abstract}
We examine the expected radio confusion on the thermal
Sunyaev-Zel'dovich (SZ) effect of galaxy clusters at 28.5 GHz and
90 GHz based on the cluster radio luminosity function (CRLF) at
1.4 GHz. With the observationally determined spectral index
(${\alpha }$, where $S_{\rm \nu }\propto \nu ^{-\alpha }$)
distribution, instead of a single average index
$\langle\alpha\rangle$, we convert the cluster radio luminosity
function at 1.4 GHz to the high frequency ones and estimate the
total radio flux in a cluster. At 28.5 GHz, radio confusion is up
to 10 $\sim 100\%$ for small clusters ($M\sim 2\times
10^{14}M_{\odot}$) below redshift $z$=1, with more severe
confusion for smaller and lower redshift clusters. By contrast, at
90 GHz the confusion is less than $10\%$ for small clusters even
at low redshifts.
\end{abstract}


\section{Introduction}
The cosmic microwave background radiation (CMBR) provides a light
source for exploration of the structure formation history. Before
the CMBR photons reach the present epoch, they interact with the
hot diffuse ionized gas residing in the galaxy clusters over a
wide range of redshifts. This process results in subtle changes to
the CMBR spectrum due to the inverse Compton scattering, the
so-called Sunyaev-Zel'dovich (SZ) effect (Sunyaev \& Zel'dovich,
1972; see Birkinshaw 1999 and Carlstrom et al. 2000 for reviews).
In the past two decades the detection of SZ effects has been
reported at an increasing rate (Grego et al. 2000; Carlstrom et
al. 2000; Joy et al. 2001; Grego et al. 2001) and this effect can
therefore provide a realistic powerful probe for cosmological and
cluster studies. Since the SZ flux is not affected by the distance
but depends only on the intrinsic thermal energy of hot gases
contained in the system, the SZ observation, in conjunction with
observations of other wavelengths, such as X-ray emitted from
intracluster gas, and optical weak lensing measurements, enables
us to determine the cosmological parameters by studying the high
redshift galaxy clusters. The SZ blank sky surveys have been
suggested (Bartlett \& Silk 1994; Barbosa et al. 1996). Several
works follow (Bartlett 2000; Holder et al. 2000; da Silva et al.
2000; Kneissl et al. 2001; Fan \& Chiueh 2001; Xue \& Wu 2001),
and have predicted how the SZ cluster counts can be related to
cosmological parameters. In addition, several next generation
interferometric arrays are either under construction or have been
proposed, e.g., AMI (Kneissl et al. 2001), SZA (Carlstrom et al.
2001), and AMiBA (Lo et al. 2001).

Radio point sources or radio galaxies are usually found in the
galaxy clusters. Therefore, the accuracy of the millimeter to
centimeter wavelength observations of the SZ effect can be
contaminated by the emission from radio sources (Loeb \& Refregier
1997). Below 218 GHz, at which the thermal SZ effect becomes null,
one measures the SZ intensity or flux decrement. The radio
sources, however, produce excess emission so that one can
underestimate the measured SZ decrement, yielding systematic
errors in the estimate of cosmological parameters. This confusion
has been estimated by Cooray et al. (1998, hereafter CGHJC), who
observed toward Abell 2218 at 28.5 GHz and concluded that the
correction to the Hubble constant is less than $6\%$. In addition
to point sources, diffuse cluster radio sources, i.e. halos and
relics, (e.g., Giovannini et al. 1993; R\"{o}ttgering et al. 1997)
also lead to confusion on the cluster SZ.

With the requirement of accurate SZ maps, the radio source
subtraction becomes essential for the low brightness (order of 100
$\mu$Jy and mJy) SZ measurements. For this reason, recent SZ
observations are performed in combination with the measurements of
radio sources (CGHJC; Komatsu et al. 1999; Pointecouteau et al.
2001). While one may remove strong sources as they can be detected
above the flux limit, the faint sources below the flux threshold
blend with the SZ signal, and result in inaccuracy. However, there
have not been systematic surveys on the radio sources in high
redshift clusters nor systematic surveys at frequencies higher
than 30 GHz. In fact, the high-frequency (e.g. 90 GHz) radio
survey at the mJy flux limit is a formidable task, as the present
telescopes are not sufficiently efficient in conducting an
unbiased survey with a wide field coverage and high sensitivity.
In view of the lack of reliable information for estimating the
radio contamination on the SZ cluster surveys at high frequencies,
such as AMiBA (90 GHz), we are motivated to derive an empirical
high-frequency cluster radio luminosity function (CRLF) by
synthesizing available low-frequency data (Ledlow \& Owen 1996,
hereafter LO96; Slee et al. 1996, hereafter SRA96, and CGHJC).
Unfortunately, the spectral evolution of radio sources is rather
uncertain. Due to the uncertainties of redshift evolution, it is
reasonable, as a first attempt, to adopt a no-evolution hypothesis
to evaluate the radio flux in high redshift clusters. In this
paper we aim to calculate the expected confusion of high-frequency
radio point sources on the cluster SZ effect over substantial
redshift and cluster mass ranges. In section 2, we briefly review
the formulation of the SZ flux. Section 3 predicts the high
frequency cluster radio luminosity functions (CRLFs) based on low
frequency surveys. The expected high frequency radio flux are
calculated in section 4. Finally, the conclusion is given in
section 5. Throughout this paper, we use $H_{\rm 0}=65$ km
s$^{-1}$ Mpc$^{-1}$, $\Omega_{\rm M}=0.35$, and $\Omega_{\rm
\Lambda }=0.65$.

\section{SZ Flux}

One may evaluate the total SZ flux by the procedure as follows (see also
Barbosa et al. 1996, Xue \& Wu 2001). For the CMB radiation passing through
the intracluster hot gas, the change in CMB intensity by thermal SZ effect
is
\begin{equation}\label{eq1}
\triangle I_{\rm \nu}=j_{\rm \nu}(\rm x)\int \left(\frac{kT_{\rm
e}}{m_{\rm e}c^{2}}\right)\sigma _{\rm T}n_{\rm e}dl,
\end{equation}
where $m_{\rm e}$ is the electron rest mass, $n_{\rm e}$ is the
electron number density, $T_{\rm e}$ is the electron temperature,
and $\sigma_{\rm T}$ is the Thomson cross section. $j_{\rm
\nu}(x)$ represents the spectral dependence
\begin{equation}\label{eq2}
j_{\rm \nu}(x)=2\frac{(kT_{\rm 0})^{3}}{(h_{\rm
p}c)^{2}}\frac{x^{4}e^{x}}{(e^{x}-1)^{2}}\left[x\coth
\left(\frac{x}{2}\right)-4\right],
\end{equation}
in which $x\equiv h_{\rm p}\nu /kT_{\rm 0}$ and $T_{\rm 0}$=2.728
K is the CMB temperature. For $\nu=90$ GHz, $x\approx$ 1.58. The
total SZ flux of a cluster at redshift $z$ is the integral of the
intensity over the solid angle subtended by the cluster
\begin{equation}\label{eq3}
S_{\rm \nu }=\frac{j_{\rm \nu }(x)}{D_{\rm A}^{2}(z)}\left( \frac{\sigma _{\rm T}}{m_{\rm e}c^{2}%
}\right) \int kT_{\rm e}n_{\rm e}dV,
\end{equation}
where $D_{\rm A}$ is the angular diameter distance to the cluster.
If we assume the gas is isothermal, then the total SZ flux depends
only on the electron number, which can be replaced by the gas mass
of the cluster. Introduce the gas fraction $f_{\rm b}=M_{\rm
gas}/M_{\rm vir}$, where $M_{\rm vir}$ is the virial mass of the
cluster, then the equation above becomes
\begin{equation}\label{eq4}
S_{\rm \nu}=\frac{j_{\rm \nu}(x)}{D_{\rm A}^{2}(z)}\left(\frac{kT_{\rm e}}{m_{\rm e}c^{2}}%
\right)\left(\frac{f_{\rm b}\sigma_{\rm T}}{\mu_{\rm e}m_{\rm
p}}\right)M_{\rm vir},
\end{equation}
where $\mu_{\rm e}=2/(1+X)$, and X=0.768 is the hydrogen mass
fraction in the primordial abundances of hydrogen and helium.

Define $R_{\rm vir}$ as the radius of the cluster within which the
mean density is $\Delta_{\rm c}$ times the critical density of the
universe $\rho_{\rm c}$ at a certain redshift, i.e., $M_{\rm
vir}=4\pi R_{vir}^3\rho_{\rm c}\Delta_{\rm c}/3$, then the
relation between temperature and mass is (Bryan \& Norman 1998)
\begin{equation}\label{eq5}
kT_{\rm e}=1.39f_{\rm T}\left(\frac{M_{\rm
vir}}{10^{15}M_{\odot}}\right) ^{2/3}\left[h^2\Delta_{\rm
c}E(z)^2\right]^{1/3}\mbox{keV},
\end{equation}
where $E(\rm z)^2=\Omega_{\rm M}(1+z)^3+\Omega_{\rm K}(1+z)^2+\Omega_{\rm \Lambda}$, $%
\Omega_{\rm K}=1-\Omega_{\rm M}-\Omega_{\rm \Lambda}$, and the
redshift-dependent Hubble constant $H(z)=100hE(z)$ km s$^{-1}$
Mpc$^{-1}$. $f_{\rm T}$ is a normalization factor obtained by
numerical simulations and about 1. $\Delta_{\rm c}$ is the density
contrast of the virialized spherical halo to the critical density
of the universe at that redshift, which can be fitted as (Eke et
al. 1996) $\Delta_{\rm
c}=18\pi^2+82[\Omega(z)-1]-39[\Omega(z)-1]^2$ for $\Omega_{\rm
K}=0$, where $\Omega(z)=\Omega_{\rm M}(1+z)^3/E^2(z)$.

The gas fraction $f_{\rm b}$ varies with the cluster temperature
by the empirical formula (Mohr et al. 1999)
\begin{equation}\label{eq6}
f_{\rm{b,50}}=(0.207\pm0.011)\left(\frac{kT}{\mbox{6
keV}}\right)^{0.34\pm0.22},
\end{equation}
where the subscript $50$ denotes $h=$0.5. We may scale $f_{\rm b}$
by $f_{\rm b}=f_{\rm{b,50}}h_{\rm 50}^{-1.5}$ (White et al. 1993),
and Eq.(\ref{eq4}) and (\ref{eq6}) will be used to compare with
the radio flux calculated in the next section, so as to determine
the radio confusion on the SZ flux. The reason for incorporating
the temperature dependence of $f_{\rm b}$ is as follows. It is
likely that preheating by supernovae and/or AGNs provides
nongravitational heating to the intracluster gas, which makes the
distribution of the hot gas extend to large radii (Wu et al. 1998,
2000). The effect is more significant in poor clusters than in
rich ones because the former have a shallow gravitational
potential, as evidenced by the steepening of the $L_{\rm x}$-$T$
relation (David et al. 1993; Wu et al. 1999) as well as the excess
entropy in galaxy clusters (Ponman et al 1999).

\section{CRLF and Radio Flux}

To estimate the total flux of the radio galaxies in clusters, we
adopt the CRLF given by LO96 (Paper VI). They did a series of
surveys on a large sample of radio galaxies in Abell clusters
using VLA, e.g., in Paper IV (Ledlow \& Owen 1995a, hereafter
LO95a), where the radio sample and cluster properties were
presented; in Paper V (Ledlow \& Owen 1995b), they did optical
observation and investigated the optical properties toward their
radio sample.

In LO96, the CRLF is for a complete sample of Abell clusters with
$z<0.09$ and can be fitted by a continuous broken power law
\begin{equation}\label{eq7}
\log{f_{\rm 1.4}}\equiv\log{\frac{N_{\rm radio}(\log{P_{\rm
1.4})}}{N_{\rm opt}}}=a+b\log{P_{\rm 1.4}},
\end{equation}
in which $N_{\rm opt}$ represents the number of galaxies over all
magnitudes brighter than $-$20.5 within 0.3 Abell radius of the
cluster center and $N_{\rm radio}(\log{P_{\rm 1.4}})$ represents
the number of radio galaxies which emit the radio power $P_{\rm
1.4}$ at 1.4 GHz. The coefficients $b$ and $a$ are fitted with the
central values as $-$0.15 and 1.77 for $P_{\rm 1.4}<10^{24.8}$
WHz$^{-1}$, and $-$1.43 and 33.67 for $P_{\rm 1.4}>10^{24.8}$
WHz$^{-1}$, respectively.

We evaluate the predicted cluster radio luminosity functions at
other frequencies, $f_{\rm \nu}$, based on the 1.4 GHz survey by
convoluting $f_{\rm 1.4}$ with the spectral index distribution
$n(\alpha)$ as
\begin{equation}\label{eq8}
\log{f_{\rm \nu}}=\int \int (\log{f_{\rm 1.4}})n(
\alpha)\delta\left[\log{P_{\rm 1.4}}-\log{P_{\rm
\nu}}-\alpha\log{\left(\frac{\rm \nu}{\mbox{1.4
GHz}}\right)}\right] d\alpha d\log{P_{\rm 1.4}},
\end{equation}
where $f_{\rm 1.4}$ is the luminosity function given by
Eq.(\ref{eq7}). The spectral index distributions $n(\alpha)$ that
we adopt are fitted by two Gaussians and normalized as
\begin{equation}\label{eq9}
n(\alpha)=\left\{
\begin{array}{ll}
0.21e^{-0.72(\alpha-1.14)^2}+1.06e^{-11.0(\alpha-0.98)^2}; & \mbox{SRA96 sample}\\
0.95e^{-8.33(\alpha-1.02)^2}+0.44e^{-3.41(\alpha-0.57)^2}; &
\mbox{CGHJC sample}
\end{array}\right.,
\end{equation}
and are shown in Figure 1. The circles are from SRA96 and the
crosses are from CGHJC. The two distributions are normalized to
their total number of sources for comparison. The SRA96 sample
contains 254 radio sources and its $n(\alpha)$ appears more
regular than that of CGHJC sample, which contains 53 sources
($\alpha$ available) only in very massive clusters
($M>10^{15}M_{\odot}$). Moreover, the spectral index in SRA96
sample is determined at lower frequencies ($1.5$ GHz $<\nu<4.9$
GHz), whereas CGHJC sample at higher frequencies ($1.4$ GHz
$<\nu<28.5$ GHz). Despite these differences, the two $n(\alpha)$
do not differ much, at least in the main body of the distribution.
We show in Figure 2 the predicted CRLFs using SRA96's $n(\alpha)$
at $\nu=28.5(1+z)$ GHz, and $90(1+z)$ GHz for $z$=0.25,
respectively. Note that the cosmological expansion has been
considered, where the frequency of the predicted function has a
redshift dependence $\nu=\nu_{\rm 0}(1+z)$, where $\nu_{\rm 0}$ is
observed frequency. For simplicity in calculating the total flux,
we fit the luminosity-weighted CRLFs by a broken power law,
\begin{equation}\label{eq10}
\log{(P_{\rm \nu}f_{\rm \nu})}=\left\{
\begin{array}{ll}
C_{\rm 1}(\nu)+\gamma_{\rm 1}(\nu)\log{P_{\rm \nu}}; & \log{P}<\log{P_{\rm b}}\\
C_{\rm 2}(\nu)+\gamma_{\rm 2}(\nu)\log{P_{\rm \nu}}; &
\log{P}>\log{P_{\rm b}}
\end{array}\right.,
\end{equation}
where $\log{P_{\rm b}}$ is the break point of a continuous
luminosity-weighted CRLF, and the fit coefficients are list in the
Table 1. We see that both the break points of SRA96 and CGHJC vary
slightly with frequency, and that the peak power of radio sources
decreases with increasing frequency. But the fit coefficients from
the two groups are not in good agreement, as shown below. This is
due to that the sample of CGHJC has flatter spectra, and the
resulting power is higher.

Assuming neither evolution effect on the spectral indices nor on
the intrinsic brightness of radio sources, we obtain the best fit
for the frequency dependence of these CRLFs coefficients,
\begin{equation}\label{eq11}
\left\{
\begin{array}{ll}
C_{\rm 1}(\nu)=&1.68+0.015\nu-2.1\times10^{-5}\nu^2 \\
\gamma_{\rm 1}(\nu)=&0.85-0.0010\nu+1.91\times10^{-6}\nu^2 \\
C_{\rm 2}(\nu)=&23.84+9.92e^{-0.015\nu} \\
\gamma_{\rm 2}(\nu)=&-0.44e^{-0.007\nu}-2.9\times10^{-4}\nu
\end{array}
\right.;\hspace{2mm}\mbox{SRA96's }n(\alpha),
\end{equation}
and
\begin{equation}\label{eq12}
\left\{
\begin{array}{ll}
C_{\rm 1}(\nu)=&2.68-0.95e^{-0.015\nu} \\
\gamma_{\rm 1}(\nu)=&0.85e^{-0.0027\nu}+1.49\times10^{-3}\nu \\
C_{\rm 2}(\nu)=&30.42+3.29e^{-0.025\nu} \\
\gamma_{\rm 2}(\nu)=&-0.43e^{-0.0022\nu}-6.14\times10^{-4}\nu
\end{array}\right.;\hspace{2mm}\mbox{CGHJC's }n(\alpha).
\end{equation}
Figure 3 shows how the coefficients depend on frequencies and
their best-fit curves. One may construct the CRLF of any redshift
by interpolating the emitted frequency $\nu$, as $\nu_{\rm
0}(1+z)$, where $\nu_{\rm 0}$ is the observed frequency.

Taking an integral over the radio power range from $10^{17}$ to
$10^{26}$ WHz$^{-1}$ and multiplying it by the total number of
cluster galaxies, we have the total radio power emitted from the
radio galaxies within a single cluster
\begin{equation}\label{eq13}
P^{\rm tot}_{\rm \nu}=N_{\rm vir}\int P_{\rm \nu}f_{\rm
\nu}dP_{\rm \nu},
\end{equation}
where $N_{\rm vir}$ is the number of galaxies in a cluster of mass
$M_{\rm vir}$. We estimate the total number of radio sources in a
cluster by the mass-to-number relation for a given cluster mass
$M_{\rm vir}$ (Carlberg et al. 1996). They use the CNOC Cluster
Survey to derive the $M_{\rm vir}/N_{\rm gal}$ ratios using the
sample with r-band absolute-magnitude limits from $M_{\rm r}^{\rm
K}=-18.0$ to $-20.0$. To be consistent with $N_{\rm opt}$ in
Eq.(\ref{eq7}), which has the optical limiting magnitude at
$-$20.5, we fit the mass-to-number relation derived by Carlberg et
al. (1996), and extrapolate the limiting magnitude to $M_{\rm
r}^{\rm K}=-20.5$ to obtain $M_{\rm vir}/N_{\rm
gal}\simeq1.5\times10^{13} h^{-1}M_{\odot}$. For a given cluster
mass $M_{\rm vir}$, we take this $N_{\rm gal}$ as the number of
galaxies $N_{\rm vir}$ in Eq.(\ref{eq13}).

To be compared with SZ flux, the total radio flux observed at
frequency $\nu_{\rm 0}$ at redshift $z$ can be evaluated from the
total power at frequency $\nu$ as
\begin{equation}\label{eq14}
S_{\rm {\nu_{\rm 0},obs}}=\frac{P^{\rm tot}_{\rm \nu}(1+z)}{4\pi
D_{\rm L}^{2}(z)},
\end{equation}
where $D_{\rm L}(z)$ is the luminosity distance to redshift $z$, $
D_{\rm L}(z)=(1+z)^{2}D_{\rm A}(z)$, and radio sources are assumed
to have no evolution. We combine Eq.(\ref{eq4}) and
Eq.(\ref{eq14}) to determine the confusion.

\section{Result}

In Figure 4, we plot the total radio flux v.s. the virial mass of
a galaxy cluster (or group) at redshift $z$=0.25, 0.5, and 1.0.
The left panels are for the observed frequency 28.5 GHz (operated
in SZA) and the right for 90 GHz (operated in AMiBA). The
$1\sigma$ error bars result primarily from the variance of the
predicted luminosity-weighted function $P_{\rm \nu}f_{\rm \nu}$,
and are much larger than the observational errors of $f_{\rm 1.4}$
and $n(\alpha)$, which are neglected. These large error bars are
due to the broad scattering of the distribution of the spectral
index $\alpha$. As a consistency check, we compared our estimate
flux at 28.5 GHz with the CGHJC sources in Figure 4, and at 1.4
GHz with the LO95a sources in Figure 5. The dash lines are our
predicted mean flux, and the data points are the radio flux
provided by LO95a. We have investigated the redshift and the X-ray
temperature of their selected clusters (c.f. Ebeling et al. 1996,
1998) and converted the cluster temperature to the virial mass by
Eq.(\ref{eq5}) to obtain the coordinates of the data points in
Figure 4 and 5. It shows that the total radio flux at 28.5 GHz of
most but few clusters in their sample are consistent with our
predictions within $1\sigma$. The same agreement is also found at
1.4 GHz. Since the measured clusters of CGHJC are massive, their
sample may have some selection bias. It may result in asymmetry in
their spectral index distribution $n(\alpha)$, leading to the
predicted radio flux higher than those from SRA96, by almost a
constant factor $\sim 2.5-3$ throughout all frequencies of
interest. It is instructive to examine the averaged spectral index
$\langle\alpha\rangle$ derived from each individual $n(\alpha)$
given by SRA96 and by CGHJC. It is found that
\begin{equation}\label{eq15}
\langle\alpha\rangle\equiv\frac{\ln[\int{(P_{\rm \nu}f_{\rm
\nu})d\log{P_{\rm \nu}}-\mbox{const.}}]}{\ln{\nu}}=\left\{
\begin{array}{ll}
-1.15; & \mbox{for SRA96}\\
-0.9; & \mbox{for CGHJC}
\end{array}\right..
\end{equation}
The $\langle\alpha\rangle$ value for CGHJC is somewhat higher than
that for SRA96, but is lower than the $\langle\alpha\rangle$
value, $-$0.77, given by CGHJC based solely on their 53 sources.
The flatter spectrum signifies the somewhat different radio
sources in massive clusters from in average clusters.

The flux ratios v.s. the cluster mass are shown in Figure 6. Note
that the SZ flux are represented in \textit{positive} values. The
flux ratio decreases with mass, since the SZ flux increases with
$M_{\rm vir}$ faster than the radio flux does. The flux ratio also
decreases with redshift under the presently adopted no-evolution
hypothesis. Moreover, at $\nu\sim$ 30 GHz, the radio flux are
higher and the SZ flux are lower than those at 90 GHz, and
therefore the flux ratios of the two frequencies differ by a large
margin. In SZ surveys for cluster mass ranging from $M\sim
2\times10^{14}M_{\odot}$ to $2\times10^{15}M_{\odot}$, the radio
contamination at 30 GHz should be a serious concern for small
clusters even at $z$=1 ($\sim 10\%$), whereas the contamination
for small clusters at 90 GHz is lower than $10\%$ even at
$z$=0.25.

\section{Conclusion}

We study the CRLFs from 1.4 GHz to 180 GHz, based on the
flux-limited 1.4 GHz observation in conjunction with studies of
radio spectral index at higher frequencies. The lack of
comprehensive radio spectra of galaxy clusters remains a problem
for predicting the radio confusion on the cluster SZ effect. In
our analysis, the cluster radio luminosity functions at high
frequencies are obtained by converting the observed low-frequency
ones with the observed spectral index distributions of radio
sources. We obtain the frequency dependence of the CRLF from two
spectral index studies. Assuming that CRLFs do not evolve, the
redshift dependence of CRLFs can be derived as well. Based on this
information, we calculate the total flux of radio point sources in
a cluster for mass ranging from $M=1\times 10^{14}M_{\odot}$ to
$5\times 10^{15}M_{\odot}$ at high frequencies and various
redshifts. Apart from few exceptional clusters, our predictions
agree with LO95a (at 1.4 GHz) and CGHJC (at 28.5 GHz) observations
in a self-consistent manner.

We also give estimates of the confusion from radio point sources
to the SZ flux. As demonstrated in Figure 6, the radio confusion
to lower frequency ($\nu\approx30$ GHz) SZ measurements poses a
severe problem for cluster of $M\lesssim 10^{15}M_{\odot}$,
whereas at higher frequency ($\nu\approx90$ GHz) the radio
confusion causes less a problem even for small clusters
($M\sim10^{14}M_{\odot}$). For observed frequencies substantially
lower than 30 GHz, such as 15 GHz proposed to be operated at AMI,
one may use Eq.(\ref{eq15}) to estimate the expected level of
radio confusion. Take $\langle\alpha\rangle=-1$; the 15 GHz radio
flux is twice higher but the SZ flux is 4 times weaker (due to the
$\nu^{2}$ Rayleigh-Jeans tail) than those in the 30 GHz
observation, and the confusion flux ratio becomes 8 times higher
than that of 30 GHz.

The estimate for the expected total confusion to a cluster of
given virial mass can be translated to the expected loss of SZ
signals, which results in reducing the cluster count and in turn
affecting the inferred values of cosmological parameters. As has
already been demonstrated, the amplitude of density fluctuations
averaged over 8$h^{-1}$ Mpc, $\sigma_{\rm 8}$, is most sensitive
to the SZ cluster count (Fan \& Chiueh 2001). The radio confusion
thus yields a lower value of the inferred $\sigma_{\rm 8}$.
Nevertheless, with the aid of a sensitive telescope for pointed
observation, the radio confusion can partially be removed by
subtraction, when the radio sources are sufficiently strong.
However weak radio sources remain to be a problem as they are
bound to blend into the SZ signals and cannot be subtracted away.
The spectral-index distribution $n(\alpha)$ can in principle
reveal the relative population of strong and weak sources at high
frequencies. But this problem can be assessed in a different way.
We have compared the observed radio flux of CGHJC (28.5 GHz) and
LO95a (1.4 GHz) with the predicted total radio flux in Figure 4
and 5, and the data show large scatters around the expected flux.
It suggests that many sources are weak and can be difficult to get
removed by subtraction. Therefore, a good SZ observing strategy is
probably to set the SZ flux limit substantially higher than the
expected flux of radio sources in a cluster. For 90 GHz
observations, a mJy flux limit is adequate for this purpose.

\acknowledgments

This work was supported in part by NSC 90-2112-M-002-026 from the
National Science Council, R.O.C. and by the National Science
Foundation of China.

\clearpage

\begin{figure}
\plotone{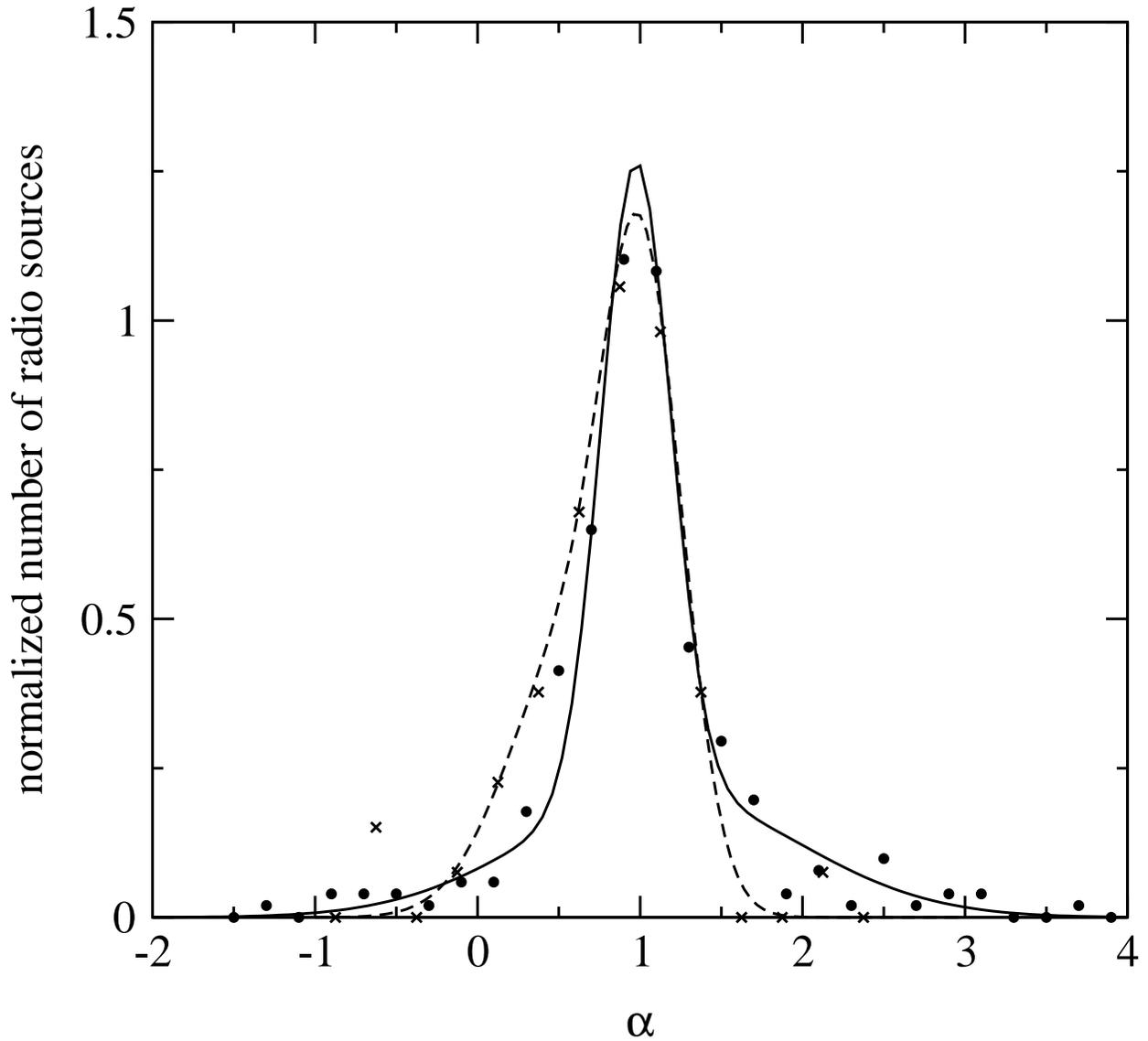} \caption{Normalized spectral index distribution
of radio sources. The circles are from SRA96 (254 sources) and the
solid line is the best fit; while the crosses are from CGHJC (53
sources) and the dash line is the best fit. CGHJC sample is biased
toward very massive clusters ($>10^{15}M_\odot$) and the spectral
index distribution is more irregular at low and high $\alpha$ than
that of SRA96.} \label{fig1}
\end{figure}

\clearpage

\begin{figure}
\plotone{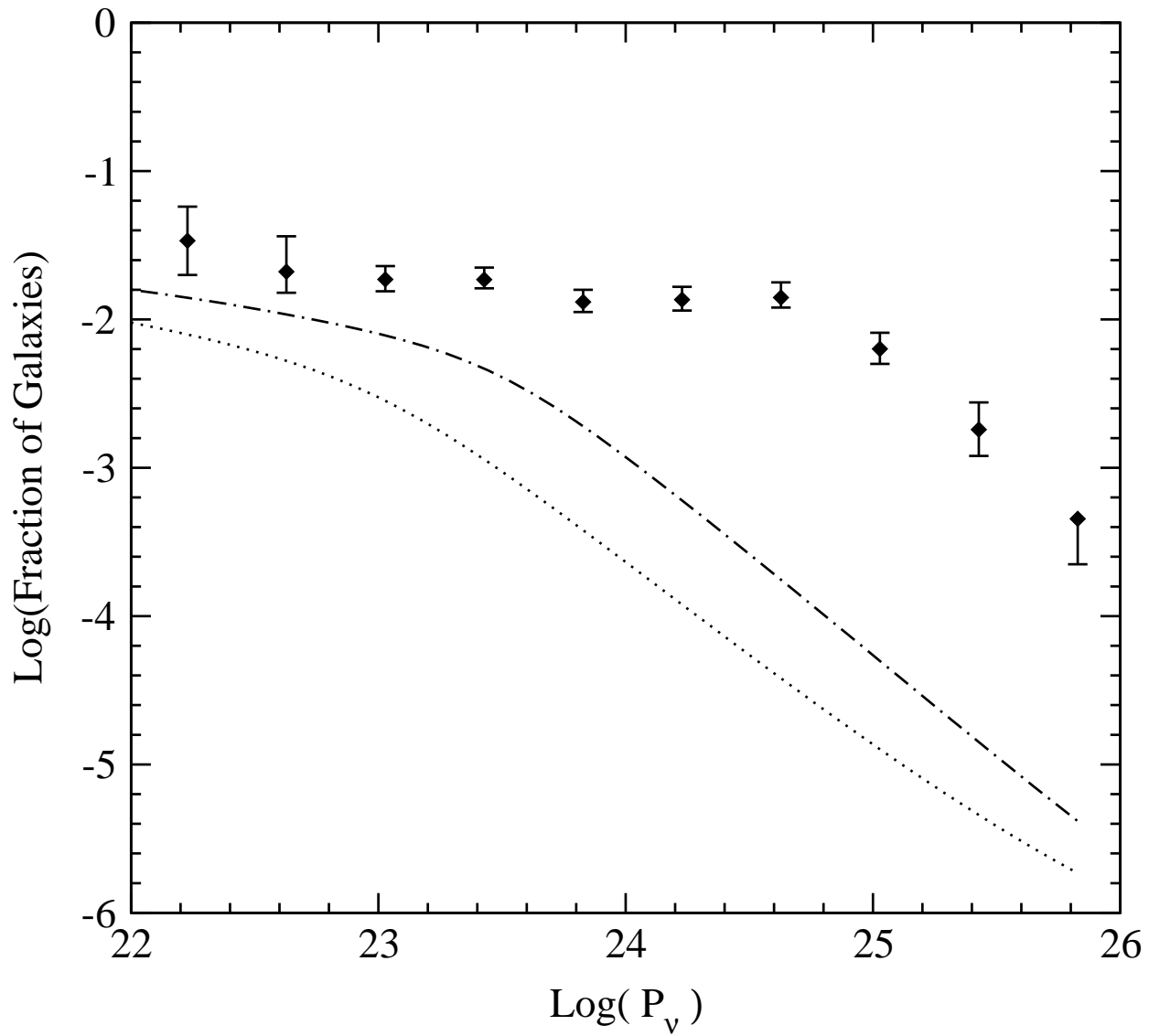} \caption{Predicted cluster radio luminosity
functions at $\nu$=28.5(1+z) GHz (the dash-dotted lines), and at
90(1+z) GHz (the dotted lines) for $z$=0.25. The data points are
at 1.4 GHz from LO96.}\label{fig2}
\end{figure}

\clearpage

\begin{figure}
\plottwo{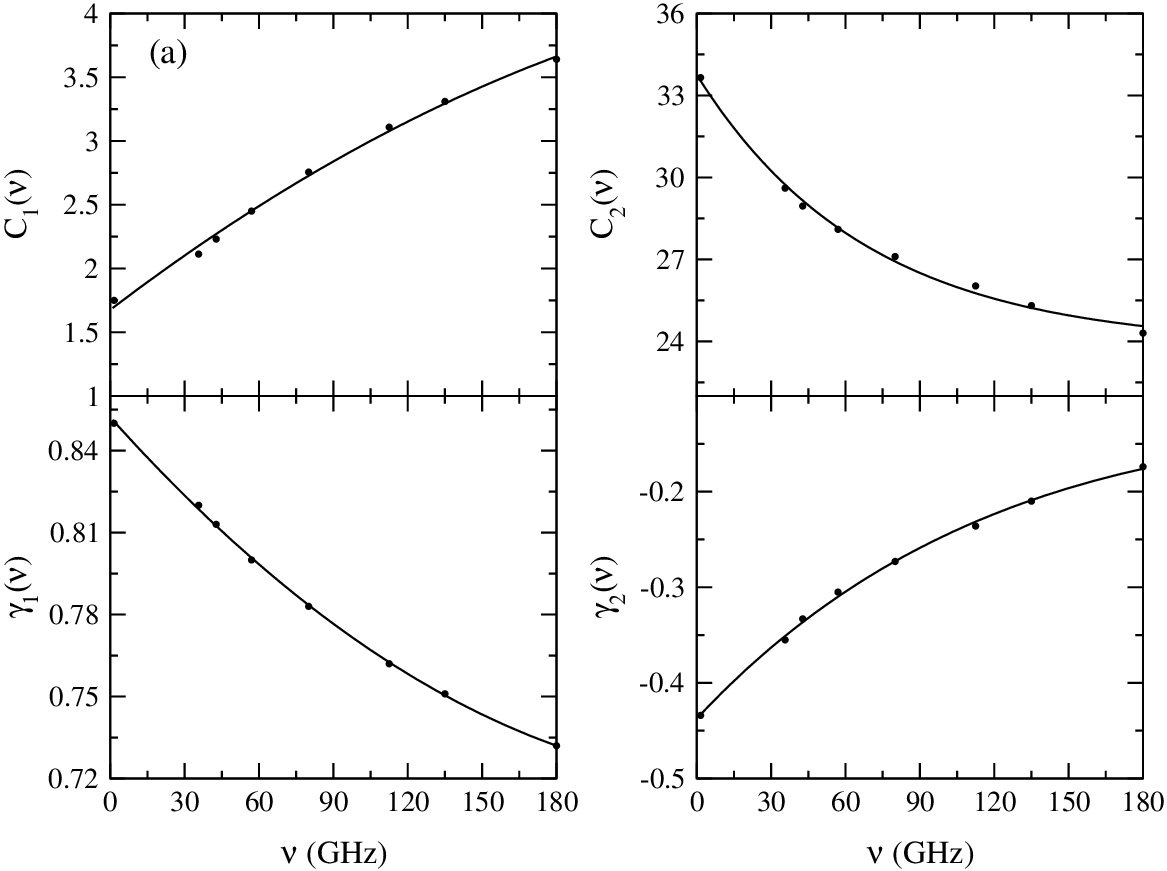}{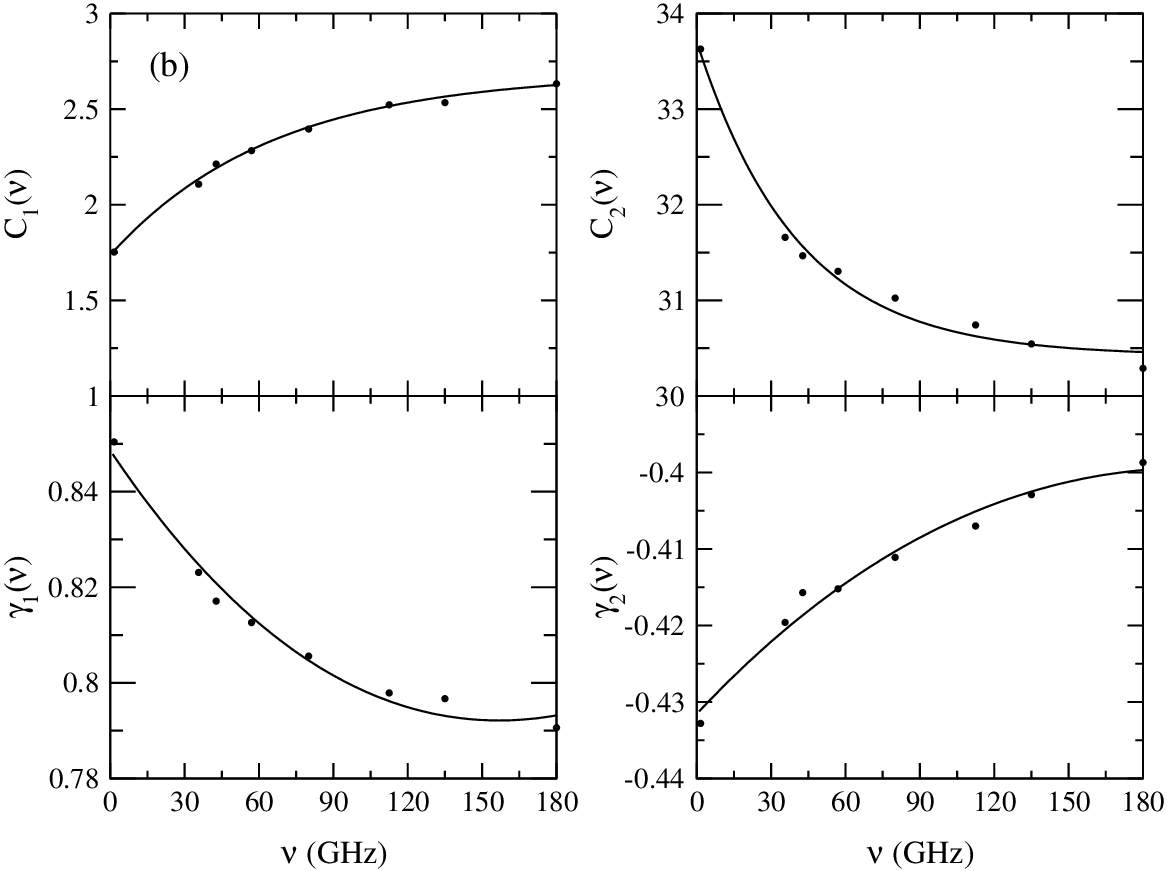} \caption{Coefficients of the
luminosity-weighted CRLFs [$\log{P_{\nu}}f_{\nu}$] from (a)SRA96's
$n(\alpha)$, and (b)CGHJC's $n(\alpha)$ in frequency.}\label{fig3}
\end{figure}

\clearpage

\begin{figure}
\plotone{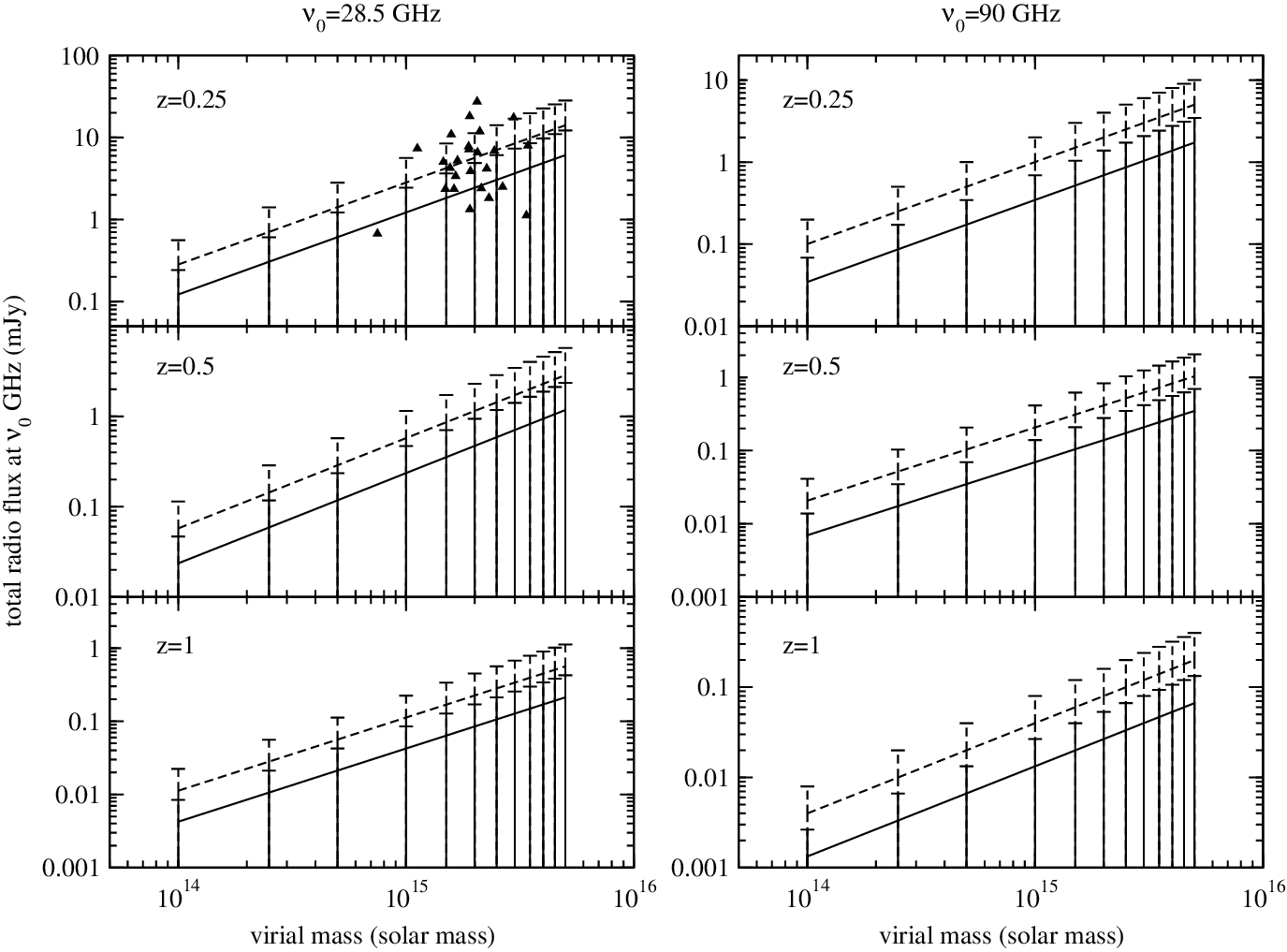} \caption{Total radio flux at $\nu$=28.5 GHz, 90
GHz in cluster virial mass with redshifts $z$=0.25, 0.5, 1.0,
respectively. The solid lines are converted from SRA96's
$n(\alpha)$, and the dash lines are from CGHJC. Triangles
represent the radio flux of 25 clusters below $z$=0.3 in
CGHJC.}\label{fig4}
\end{figure}

\clearpage

\begin{figure}
\plotone{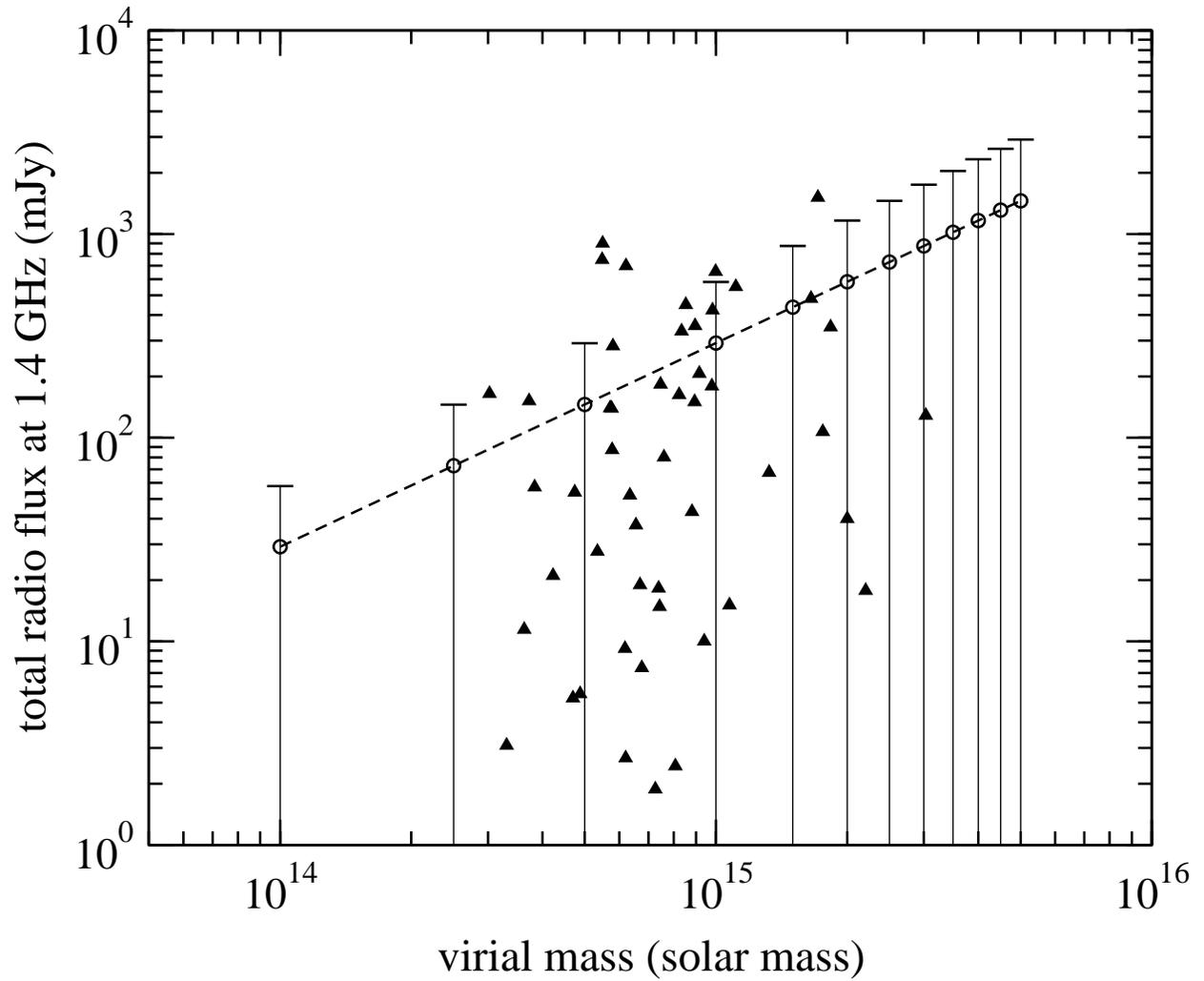} \caption{Total radio flux at 1.4 GHz from our
predicted CRLFs. The data points are boosted up to $z$=0.09 from
LO95a.}\label{fig5}
\end{figure}

\clearpage

\begin{figure}
\plotone{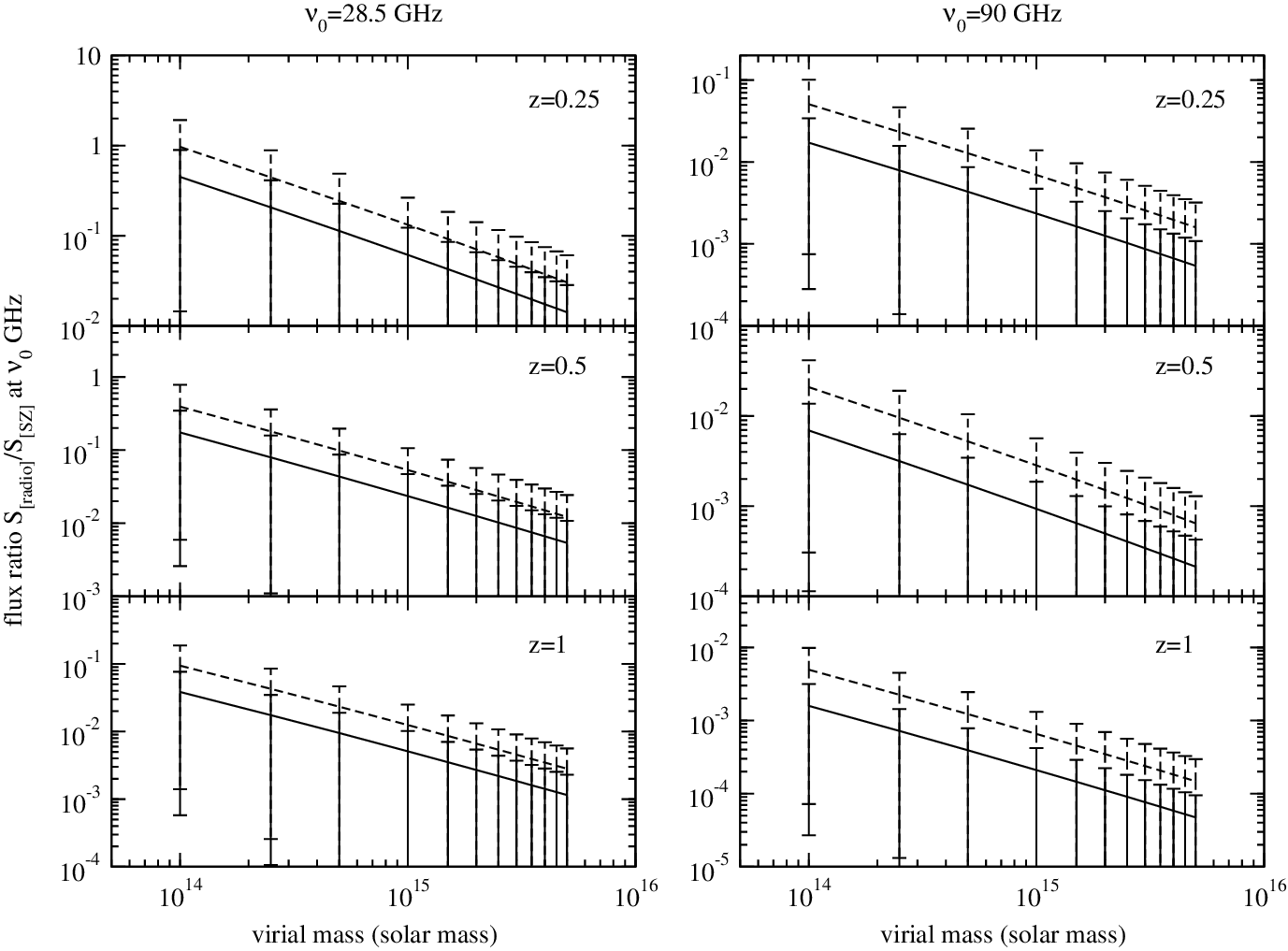} \caption{Flux ratios at $\nu$=28.5 GHz, 90 GHz in
cluster virial mass with redshift $z$=0.25, 0.5, 1.0,
respectively. The curves are the same as the former.}\label{fig6}
\end{figure}

\clearpage

\begin{table}
\caption{Fit parameters for the dependence of the
luminosity-weighted CRLFs [$\log{P_{\nu}f_{\nu}}$] from
$n(\alpha)$ in SRA96 and CGHJC, respectively.}\label{tb1}\vskip
0.1truein
\begin{tabular}{cccccccccc}
\tableline \tableline
  Author & $\nu_0$ GHz & $z$ & $\nu$ GHz & $C_1$ & $\gamma_1$ & $C_2$ & $\gamma_2$ & $\log{P_b}$\\
\tableline
 SRA96 & 1.4 & 0.09 & 1.526 & 1.75 & 0.85 & 33.65 & $-$0.43 & 24.8 \\
   & 28.5 & 0.25 & 35.625 & 2.11 & 0.82 & 29.61 & $-$0.36 & 23.4 \\
   & 28.5 & 0.5 & 42.75 & 2.23 & 0.81 & 28.95 & $-$0.33 & 23.3 \\
   & 28.5 & 1.0 & 57.0 & 2.45 & 0.80 & 28.10 & $-$0.31 & 23.2 \\
   & 40.0 & 1.0 & 80.0 & 2.76 & 0.78 & 27.11 & $-$0.27 & 23.1 \\
   & 90.0 & 0.25 & 112.5 & 3.11 & 0.76 & 26.03 & $-$0.24 & 23.0 \\
   & 90.0 & 0.5 & 135.0 & 3.31 & 0.75 & 25.31 & $-$0.21 & 22.9 \\
   & 90.0 & 1.0 & 180.0 & 3.64 & 0.73 & 24.30 & $-$0.17 & 22.8 \\
\tableline
 CGHJC & 1.4 & 0.09 & 1.526 & 1.75 & 0.85 & 33.63 & $-$0.43 & 24.8 \\
   & 28.5 & 0.25 & 35.625 & 2.11 & 0.82 & 31.66 & $-$0.42 & 23.8 \\
   & 28.5 & 0.5 & 42.75 & 2.21 & 0.82 & 31.47 & $-$0.42 & 23.7 \\
   & 28.5 & 1.0 & 57.0 & 2.28 & 0.81 & 31.30 & $-$0.42 & 23.6 \\
   & 40.0 & 1.0 & 80.0 & 2.40 & 0.81 & 31.02 & $-$0.41 & 23.5 \\
   & 90.0 & 0.25 & 112.5 & 2.52 & 0.80 & 30.74 & $-$0.41 & 23.4 \\
   & 90.0 & 0.5 & 135.0 & 2.53 & 0.80 & 30.55 & $-$0.40 & 23.3 \\
   & 90.0 & 1.0 & 180.0 & 2.63 & 0.79 & 30.29 & $-$0.40 & 23.2 \\
\tableline
\end{tabular}
\end{table}

\end{document}